\documentclass[10pt, conference, compsocconf]{IEEEtran}
\usepackage[dvips]{graphicx}
\ifCLASSINFOpdf
\else
\fi
\begin{document}
%
\title{Adaptive Fractal-like Network Structure for Efficient Search\\
of Inhomogeneously Distributed Targets at Unknown Positions
$^{1}$}

\author{\IEEEauthorblockN{Yukio Hayashi (yhayashi@jaist.ac.jp)}
\IEEEauthorblockA{
Japan Advanced Institute of Science and Technology}
}


%


\maketitle

\begin{abstract}
Since a spatial distribution of communication requests is inhomogeneous
and related to a population,
in constructing a network, it is crucial for
delivering packets on short paths through the links
between proximity nodes and for
distributing the load of nodes
how to locate the nodes as base-stations
on a realistic wireless environment.
In this paper, from viewpoints of complex network science and
biological foraging,
we propose a scalably self-organized geographical network,
in which the proper positions of nodes and the network topology are
simultaneously determined according to the population,
by iterative divisions of rectangles for load balancing of nodes
in the adaptive change of their territories.
In particular,
we consider a decentralized routing by using only local information,
and show that, for searching targets around high population areas,
the routing on the naturally embedded fractal-like structure by population
has higher efficiency
than the conventionally optimal strategy on a square lattice.
\end{abstract}

\begin{IEEEkeywords}
self-organised design;
wide-area wireless communication; 
routing in ad hoc networks;
random walk; Levy flight
\end{IEEEkeywords}

%
\IEEEpeerreviewmaketitle

\section{Introduction}
\footnotetext[1]{ADAPTIVE2012, July 22-27, Nice, France, 

Copyright\copyright by IARIA}

Many network infrastructures: 
power grids, airline networks, and the Internet,
are embedded in a metric space,
and long-range links are relatively restricted  \cite{Yook02,Gastner06}
for economical reasons.
The spatial distribution of nodes is neither
uniformly at random nor on a regular lattice,
which is often assumed in the conventional network models.
In real data,
a population density is mapped to the number of
router nodes on Earth \cite{Yook02}.
Similar spatially inhomogeneous distributions of nodes are
found in air transportation networks \cite{Guimera05}
and in mobile communication networks \cite{Lambiotte08}.
Thus, it is not trivial how to locate nodes on a space in
a pattern formation of points.
Point processes in spatial statistics \cite{Stoyan95}
provide models for irregular patterns
of points in urban planing, astronomy, forestry, or ecology, such as
spatial distributions of rainfall, germinations, plants, and animals.
The processes assume
homogeneous Poisson and Gibbs distributions to generate a pattern of
random packing or independent clustering, and to estimate parameters of
competitive potential functions in a territory model
for a given statistical data, respectively.
However, rather than random pattern and statistical estimation,
{\bf we focus on
a self-organized network infrastructure 
by taking into account realistic
spatial distributions of nodes and communication requests}.
In particular, we aim to develop adaptive and 
scalable ad hoc networks by adding the links between proximity nodes 
according to the increasing of communication requests.
Because a spatial distribution of communication requests affect the
proper positions of nodes,
which control both the load of requests assigned to each node
(e.g., assigned at the nearest access point of node 
as a base-station from a user)
and the communication efficiency depending on the selection of
routing paths.

For a routing in ad hoc networks, global information,
e.g., a routing table in the Internet, cannot be applied,
because many nodes and connections between them
are likely to change over time.
In early works on computer science,
some decentralized routing methods were developed
to reduce energy consumption in sensor or mobile networks.
However they lead to the failure of
guaranteed delivery \cite{Urrutia02};
in the flooding algorithm, multiple redundant
copies of a message are sent and cause congestion, while
greedy and compass routings may occasionally fall into infinite loops
or into a dead end. 
In complex network science,
other efficient decentralized routing methods
have been also proposed.
The stochastic methods by using local information 
of the node degrees and other measures
are called preferential \cite{Wang06b}
and congestion-aware \cite{Danila06,Wang06c} routings 
as extensions of a uniformly random walk.

Decentralized routing 
has a potential performance to search
a target whose position is unknown in advance.
Since this situation looks like foraging,
the biological strategy may be useful for the efficient search.
{\bf We are interested in a relation of the search 
and the routing on a spatially inhomogeneous 
network structure according to a population}.
Many experimental observations for biological foraging
found the evidence in favor of anomalous diffusion
in the movement of insects,
fishes, birds, mammals and human being \cite{Viswanathan11}.
As the consistent result,
it has been theoretically analyzed for a continuous space model
that an inverse square root distribution of flight lengths
is an optimal strategy to search sparsely and randomly
located targets on a homegeneous space \cite{Viswanathan99}.
The discrete space models on a regular lattice \cite{Santos05}
and the defective one \cite{Santos08} are also discussed.
Such behavior
is called {\it Levy flight} characterized by a distribution function
$P(l_{ij}) \sim l_{ij}^{- \mu}$ with $1 < \mu \leq 3$,
where $l_{ij}$ is a flight length between nodes $i$ and $j$
in the stochastic movement for any direction.
The values of $\mu \ge 3$ lead to Brownian motions,
while $\mu \rightarrow 1$ to ballistic motions.
The optimal case is $\mu \approx 2$
for maximizing the efficiency of search.
Here, we assume that the mobility of a node is 
ignored due to a sufficiently slow speed in comparison with 
the communication process.
In the current or future technologies, 
wide-area wireless connections by directional beams will be possible, 
the modeling of unit disk graph with a constant transmission range 
is not necessary.
Thus,
{\bf we propose a scalably self-organized geographical network  
and show that the naturally embedded fractal-like structure is 
suitable for searching targets more efficiently 
than the square lattice tracked by the Levy flights}.

\section{Geograhical Networks}
\subsection{Conventional Models}
Geographical constructions of complex networks have been 
proposed so far.
As a typical generation mechanism of scale-free (SF) networks
that follow a power law degree distribution
found in many real systems \cite{Albert02,Newman03},
a spatially preferential attachment is applied in
some extensions
\cite{Brunet02,Manna02,Manna03,Nandi07,Wang09}
from the topological degree based model \cite{Barabasi99}
to a combination of degree and distance based model.
The original preferential attachment is known as
``rich gets richer'' rule that means a higher degree node tend to get
more links.
On the other hand,
geometric construction methods have also been proposed.
They have both small-world \cite{Watts98} 
and SF structures generated by a recursive growing
rule for the division of a chosen triangle
\cite{Zhang08,Zhou05,Zhang06,Doye05}
or for the attachment aiming at a chosen edge
\cite{Wang06a,Rozenfeld07,Dorogovtsev02}
in random or hierarchical selections.
These models are proper
for the analysis of degree distribution
due to the regularly recursive generation process.
Although
the position of a newly added node is basically free as far as the
geometric operations are possible,
it has no relation to population.
Considering the effects of population on a geographical
network is necessary to self-organize a spatial
distribution of nodes
that is suitable for socioeconomic communication
and transportation requests.
Moreover, in these geometric methods,
narrow triangles with long links tend to be constructed,
and adding only one node per step may lead to
exclude other topologies from the SF structure.
Unfortunately, 
SF networks are extremely vulnerable against the
intentional hub attacks \cite{Albert00}.
We should develop other self-organizations of network
apart from the conventional models;
e.g. a better network without long links
can be constructed by subdivisions of equilateral triangle
which is a well balanced (neither fat nor thin)
shape for any directions.

\subsection{Generalized Multi-Scale Quartered Network}
Thus, 
we have considered the multi-scale quartered (MSQ) network model 
\cite{Hayashi09,Hayashi10}. 
It is based on a stochastic construction
by a self-similar tiling of primitive shape,
such as an equilateral triangle or square.
The MSQ networks have several advantages of the strong robustness of
connectivity against node removals by random failures and intentional
attacks, the bounded short path as $t=2$-spanner
\cite{Karavelas01}, and the
efficient face routing by using only local information.
Furthermore, the MSQ networks 
are more efficient (economic) with shorter link lengths
and more suitable (tolerant)
with lower load for avoiding traffic congestion \cite{Hayashi10}
than the state-of-the-art geometric growing networks
\cite{Zhang08,Wang06a,Rozenfeld07,Dorogovtsev02,Zhou05,Zhang06,Doye05}
and the spatially preferential attachment models 
\cite{Brunet02,Manna02,Manna03,Nandi07,Wang09} 
with various topologies
ranging from river to SF geographical networks. 
However, in the MSQ networks, the position of a new node is restricted
on the half-point of an edge of the chosen face, and the
link length is proportional to $(\frac{1}{2})^{H}$
where $H$ is the depth number of iterative divisions.
Thus, from square to rectangle, we generalize the division procedures
as follows.
\begin{description}
  \item[Step0: ] Set an initial square whose inside
             are the candidates of division axes as
             the segments of a $L \times L$ lattice.
  \item[Step1: ]  At each time step, a face is chosen with a
             probability proportional to the population counted in the
             face covered by mesh blocks of a census data.
  \item[Step2: ] Four smaller rectangles are created from the
             division of the chosen rectangle face by horizontal
             and vertical axes.
             For the division, two axes are chosen by that
             their cross point is the nearest to
             the population barycenter of the face.
  \item[Step3: ] Return to Step 1, while the network size 
	     (the total number of nodes) $N$ does not exceed a given size.
\end{description}
Note that the maximum size $N_{max}$ depends on the value of $L$;
the iteration of division is finitely stopped,
since the extreme rectangle can not be divided any longer
when one of the edge lengths of rectangle is the initial
lattice's unit length.
We use the population data on a map in $80 km^{2}$ of 
$160 \times 160$ mesh blocks ($L = 160$) provided by the 
Japan Statistical Association.
Of course, other date is possible.

It is worth noting that
the positions of nodes and the network topology are simultaneously
determined by the divisions of faces
within the fractal-like structure.
There exists a mixture of sparse and dense parts of nodes
with small and large faces.
Moreover, 
with the growing network, 
the divisions of faces perform a load balancing of nodes 
in their adaptively changed territories for the population.
We emphasize that 
such a network is constructed according to a spatially inhomogeneous 
distribution of population which is  proportional to
communication requests in a realistic environment.
In the following, we show 
the naturally embedded fractal-like structure are
suitable for searching targets.

\section{Search Performance}
\begin{figure}[htb]

\hspace{-1cm}
 \includegraphics[height=73mm]{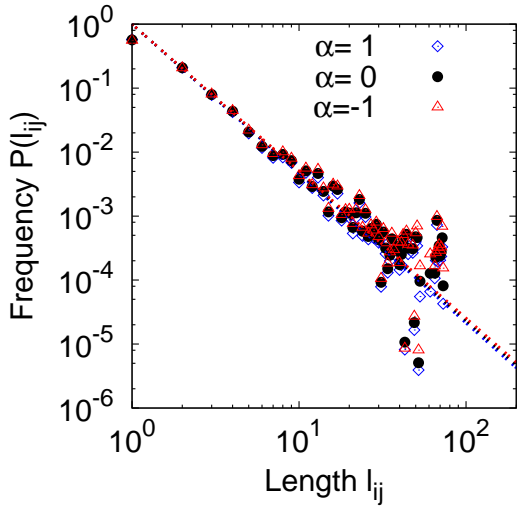}
\caption{Length distribution of 
visited links on generalized MSQ networks
by an $\alpha$-random walker in $10^{6}$ time steps.
The marks of blue diamond, black circle, and red triangle
correspond to the cases of $\alpha = 1, 0, -1$, respectively.
These results are obtained by the average of 100 networks
for $N = 2000$.}
\label{fig_vis_freq_link}
\end{figure}

\begin{figure}[htb]
\centering
 \includegraphics[height=100mm]{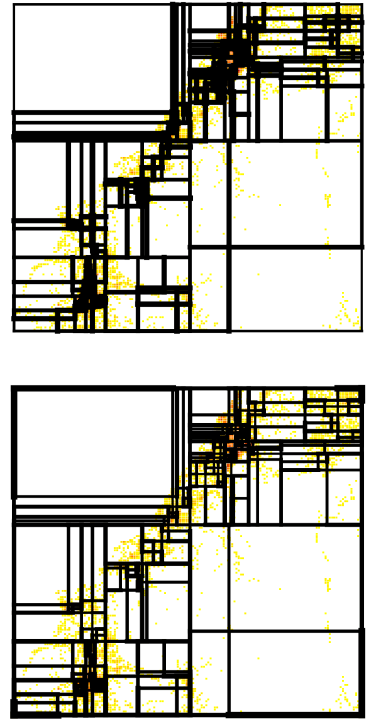}
\caption{Visualization examples of the visited links 
by $\alpha$-random walks
at $\alpha = 1$ (Top) and $\alpha = -1$ (Bottom)
on a generalized MSQ network for $N = 500$.
The thickness of link indicates the frequency of
visiting in $10^{6}$ time steps.
From light to dark: white, yellow, and orange to red,
the color gradation on a mesh block is proportionally assigned to
the population.
Many nodes represented as cross points of links
concentrate on high population
(dark: orange and red) areas on the diagonal direction.
In the upper left and lower right of square, 
corner triangle areas 
lighted by almost white are the sea of Japan and the Hakusan 
mountain range.}
\label{fig_visualization}
\end{figure}

As a preliminary, 
we consider the preferential routing \cite{Wang06b} 
which is also called $\alpha$-random walk \cite{Hayashi11}; 
The forwarding node $j$ is chosen proportionally to
$K_{j}^{\alpha}$ by a walker
in the connected one hop neighbors ${\cal N}_{i}$
of its resident node $i$ of a walker (packet), 
where $K_{j}$ denotes the degree of node $j$ and 
$\alpha$ is a real parameter.
We assume that the start position of walker
is set to the nearest node to the population barycenter 
of the initial square.
Figure \ref{fig_vis_freq_link} shows the length distribution of 
visited links.
The dashed lines in log-log plot imply a power law,
for which the exponents 
estimated as the slopes by a mean-square-error method 
are  2.336, 2.315, and 2.296 for $\alpha = 1, 0, -1$, respectively.
These values are close to the optimal exponent 
$\mu \approx 2$ \cite{Viswanathan99,Santos05} in the Levy flight
on a square lattice. 
The exponents for the $\alpha$-random walks slightly increase
as the network size $N$ becomes larger.
Here, the case of $\alpha = 0$ shows the distribution
of existing links on a network.
Since the stationary probability of incoming at node $j$
is $P_{j}^{\infty} \propto K_{j}^{1+\alpha}$
\cite{Noh04}, 
especially at $\alpha = 0$, each of the connected links to $j$
is chosen at random
by the probability $1/K_{j}$ for the leaving from $j$,
therefore a walker visit each link at the same number.
Figure \ref{fig_visualization} shows that 
the frequency of visited links by the $\alpha$-random walks 
at $\alpha = \pm 1$ is different even for the degrees 3 and 4 
in a generalized MSQ network.
On the thick lines,
a walker tends to visit high population (diagonal) areas
colored by orange and red in the case of $\alpha = 1$,
while it tends to visit low population
peripheral (corner) areas in the case of $\alpha = -1$.
Thus, the case of $\alpha = 1$ is expected
to selectively cover high population areas which has a lot of
communication requests in cities.
Note that the absolute value of $\alpha$ should be not too large,
since a walker is trapped between high/low degree nodes in a
long time as the ping-pong phenomena which does not contribute
to the search of targets.

\begin{figure}[h]
\centering
\includegraphics[height=40mm]{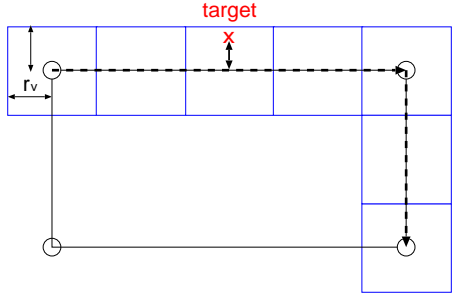}
\caption{Searching in a generalized MSQ network.
Each blue square represents a vision area,
and is scanned (from left to right,
from top to bottom in this example)
by the walker on an edge
between two nodes (denoted by circles) of a rectangle.
For a target in the area, the walker moves to get it
and returns on the link.}
\label{fig_scan_search}
\end{figure}

We investigate the search efficiency for the $\alpha$-random
walk on a generalized MSQ network,
and compare the efficiency with that for
Levy flights on a $L \times L$ square lattice
with periodic boundary conditions \cite{Santos05}.
As shown in Fig. \ref{fig_scan_search},
a walker constantly looks for
targets (destination nodes of packets) scanning
on a link between two nodes in the generalized MSQ network.
If a target exists in the vision area of $r_{v}$ hops for the
up/down/left/right directions from the center position,
a walker gets it
and return to the position on the link
for continuing the search on the same direction. 
When more than one targets exit in the area, 
a walker gets all of them successively in
each direction, and return to the position. 
Only at a node of rectangle, 
the search direction is changeable along
one of the connected links.
Thus, the search is restricted
on the edges of rectangle in the generalized MSQ network.
While the search direction of a Levy flight 
on the square lattice \cite{Santos05} is 
selectable from four directions of horizontal and vertical
at all times after getting a target 
in the scanning with the vision area of $r_{v}$ hops, 
moreover, 
the length of scan follows 
$P(l_{ij}) \sim l_{ij}^{-\mu}$, $l_{ij}> r_{v}$.
We set a target at the position chosen
proportionally to the population around a cross point in $(L+1)^{2}$, 
for which the population is defined by the average of
four values in its contact mesh regions.
In particular, we discuss the destructive case \cite{Santos05}: 
once a target is detected by a walker, then it is removed and a new
target is created at a different position chosen with the above
probability.

The search efficiency \cite{Viswanathan99,Santos05,Santos08}
is defined by
\begin{equation}
  \eta \stackrel{\rm def}{=}
   \frac{1}{M} \sum_{m=1}^{M} \frac{N_{s}}{L_{m}},
\label{eq_eta}
\end{equation}
\begin{equation}
  \lambda \stackrel{\rm def}{=}
   \frac{(L+1)^{2}}{N_{t} 2 r_{v}},
\label{eq_lambda}
\end{equation}
where $L_{m}$ denotes the traversed distance counted by
the lattice's unit length until detecting
$N_{s} = 50$ targets from the total 
$N_{t}$ targets in the $m$th run.
We consider a variety of 
$N_{t} = 60, 100, 200, 300, 400$, and $500$ 
for investigating the dependency of the search efficiency 
on the number $N_{t}$ of targets.
The quantity $\lambda$ represents the
mean interval between two targets 
for the scaling of efficiency by target density.
We set 
$M = 10^{3}$ and $r_{v} = 1$ for the convenience of simulation.
Intuitively,  the sparse and dense structures according to the 
network size $N$ have the advantage and disadvantage 
in order to raise the search efficiency in the generalized MSQ network. 
Although the scanned areas are limited by some large rectangle holes 
as $N$ is small,
a walker preferably  visits the high population areas
which include many targets.
While the scanned areas are densely covered as $N$ is large,
the search direction is constrained on long links
of a collapse rectangle, therefore it is rather hard for a walker
to escape from a local area in which targets are a few.

\begin{figure}[htb]
\centering
 \includegraphics[height=100mm]{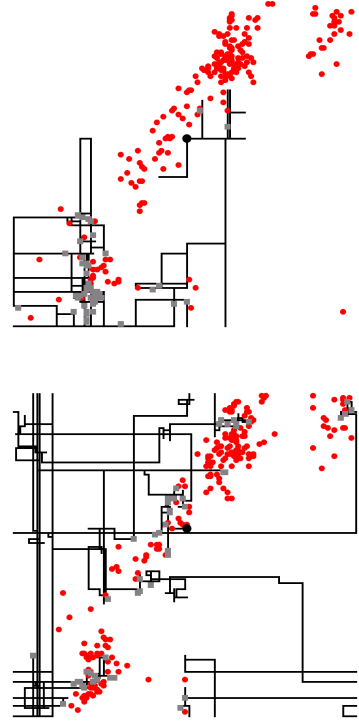}
\caption{Trajectories of a random walk (Top)
at $\alpha = 0$ on a generalized MSQ network for $N = 500$ 
and of a Levy flight (Bottom) for $\mu = 1.8$ on the square lattice 
with periodic boundary conditions 
until detecting $N_{s} = 50$ targets in $N_{t} = 200$.
Black circle, red circles, and gray rectangle marks denote 
the start point at the population barycenter, 
the existing targets, 
and the removed targets after the detections, respectively.
Note that a walker can travel 
back and forth on a link in the connected path.}
\label{fig_traject}
\end{figure}

\begin{figure}[htb]
  \begin{minipage}[htb]{.47\textwidth} \hspace{-1cm}
    \includegraphics[height=62mm]{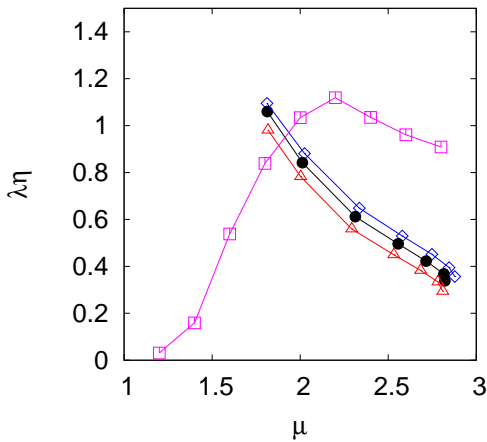}
    \begin{center} (a) $N_{t} = 100$ \end{center}
  \end{minipage}
  \hfill
  \begin{minipage}[htb]{.47\textwidth} \hspace{-1cm}
    \includegraphics[height=62mm]{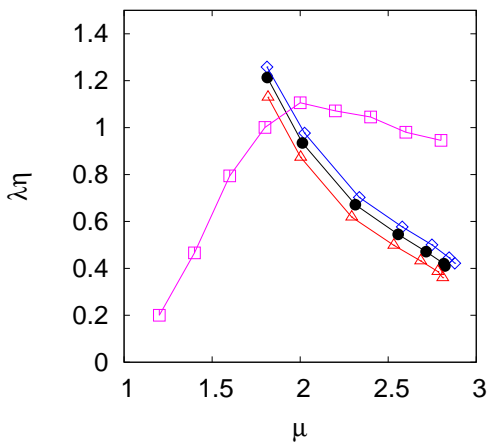}
    \begin{center} (b) $N_{t} = 200$ \end{center}
  \end{minipage}
\caption{The scaled efficiency $\lambda\eta$ vs. the exponent
$\mu$.
The marks of blue diamond, black circle, and red triangle
correspond to the cases of $\alpha = 1, 0, -1$, respectively,
in which the increasing values of $\mu$ are estimated
for generalized MSQ networks at 
$N = 500, 1000, 2000, 3000, 4000, 5000$,
and $5649$: $N_{max}$ from left to right.
The magenta rectangle corresponds to the case of Levy flights
on the square lattice.
These results are obtained by the average of 100 networks.}
\label{fig_efficiency}
\end{figure}

We compare the search efficiency of $\alpha$-random walks 
in the generalized MSQ networks
with that of the Levy flights in the square lattice.
Figure \ref{fig_traject} shows typical trajectories 
until detecting $N_{s} = 50$ targets.
On the generalized MSQ network and the square lattice, 
a walker tends to cover a local area with high population 
and a wider area, respectively.
Without wandering peripheral wasteful areas, 
the generalized MSQ network has a more efficient structure 
than the square lattice 
for detecting many targets concentrated on the 
diagonal areas.
Here the exponent $\mu = 1.8$ of Levy flight 
corresponds to the slope of $P(l_{ij})$ 
in the generalized MSQ network at the optimal size $N = 500$
for the search efficiency. 
As shown in Fig. \ref{fig_efficiency}(a)(b),
the generalized MSQ networks of $N = 500$ 
(the diamond, circle, and triangle marks are sticking out at the left)
have higher efficiency than the square lattice 
(the rectangle mark).
For the cases with many nodes of $N \geq 1000$,
the efficiency is decreased more rapidly than that of the Levy flight,
however this phenomenon means that 
many nodes are wasteful and unnecessary to get a high search performance 
in generalized MSQ networks.
When the number $N_{t}$ of targets increases in cases 
from Fig. \ref{fig_efficiency}(a) to (b), 
the curves are shift up, 
especially for the generalized MSQ networks.
The peak value for $N_{t} = 200$ is larger than the optimal case of
the Levy flight at $\mu = 2.0$.
Therefore denser targets to that extent around $N_{t} = 200$ is suitable, 
although a case of larger $N_{t} > 300$ 
brings down the search efficiency 
even for inhomogeneously distributed targets.

\begin{figure}[htb]

\hspace{-1cm}
 \includegraphics[height=70mm]{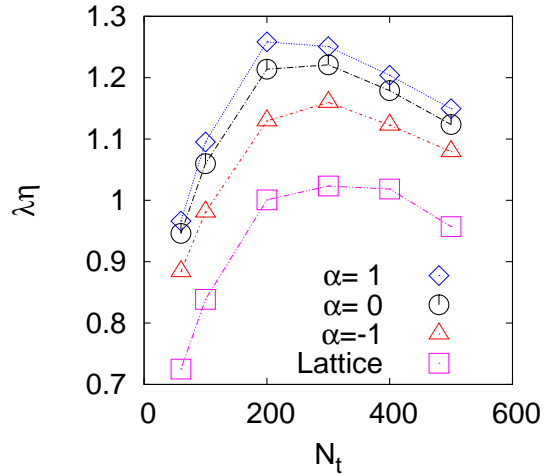} 
\caption{The number $N_{t}$ of targets vs. the scaled efficiency 
$\lambda \eta$ of 
$\alpha$-random walks on the generalized MSQ networks for $N=500$
and of the corresponding 
Levy flights for $\mu = 1.8$ (see Fig. \ref{fig_efficiency})
on the square lattice.
The maximum (optimal) efficiency appears in $N_{t} = 200 \sim 300$.
These results are obtained by the average of 100 networks.}
\label{fig_Nt_vs_efficiency}
\end{figure}

In more details,
Fig. \ref{fig_Nt_vs_efficiency} shows the effect of 
the number $N_{t}$ of targets 
on the search efficiency $\lambda \eta$.
The efficiency firstly increases, 
then reaches at a peak, and finally decreases for 
setting more targets.
This up-down phenomenon is caused from a trade-off between 
$L_{m}$ and $N_{t}$ in Eqs. (\ref{eq_eta}) and (\ref{eq_lambda}).
Note that 
the case of size $N < 500$ is omitted for the generalized MSQ networks.
Because sometimes the process for detecting targets until $N_{s}$ 
is not completed, moreover, 
the variety of link lengths is too little to estimate the
exponent as a slope of $P(l_{ij})$ in the log-log scale.         
In other words, the estimation is
inaccurate because of the short linear part.

\section{Conclusion}
We have considered a scalably self-organized geographical network 
by iterative divisions of rectangles for load balancing of 
nodes in the adaptive change of their territories according to 
the increasing of communication requests.
In particular, 
the spatially inhomogeneous distributions of population 
and the corresponding positions of nodes are important.
For the proposed networks, 
we have investigates the search efficiency 
in the destructive case \cite{Santos05} with new creations of target
after the detections, and shown that 
the $\alpha$-random walks as decentralized routing on the networks 
have higher search efficiency 
than the Levy flights known as the optimal strategy 
\cite{Viswanathan99,Santos05} 
on the square lattice with periodic boundary conditions.
{\bf One reason for the better performance is 
the anisotropic covering of high population areas}.
Thus, 
the naturally embedded fractal-like structure is suitable for 
searching targets in such a realistic situation.
In more rigorous discussions about the performance, 
statistical tests \cite{Clauset09}
may be useful to clarify the applicability of the proposed method.


\section*{Acknowledgment}
The author would like to thank Mitsugu Matsushita (Chuo University)
for his valuable comments
and Takayuki Komaki (JAIST) for helping the simulation.
This research is supported in part by
Grant-in-Aide for Scientific Research in Japan, No.21500072.



%

\end{document}